\documentclass[aps,prl,twocolumn,superscriptaddress,showpacs,floatfix]{revtex4-2}
\usepackage{graphicx,bm,epsfig,textcomp,color,dcolumn,setspace,array,mathrsfs,amsmath,amssymb,gensymb,booktabs,url,hyperref,bibentry,natbib}
\begin{document}

\title{Strain-engineered long-distance supercurrent in an altermagnetic CrSb based Josephson junction}

\author{Yifan Gao}
\thanks{These authors contributed equally to this work.}
\affiliation{Frontier Institute of Chip and System, State Key Laboratory of Integrated Chips and Systems, Zhangjiang Fudan International Innovation Center, Fudan University, Shanghai 200433, China}

\author{Jin Yan}
\thanks{These authors contributed equally to this work.}
\affiliation{Shanghai Institute of Microsystem and Information Technology, ShanghaiTech University, Shanghai 201210, China}

\author{Junjie Wei}
\thanks{These authors contributed equally to this work.}
\affiliation{Frontier Institute of Chip and System, State Key Laboratory of Integrated Chips and Systems, Zhangjiang Fudan International Innovation Center, Fudan University, Shanghai 200433, China}

\author{Yifan Zhang}
\affiliation{Shanghai Institute of Microsystem and Information Technology, ShanghaiTech University, Shanghai 201210, China}

\author{Zhiyuan Zhou}
\affiliation{Key Laboratory of Advanced Materials (MOE), School of Materials Science and Engineering, Tsinghua University, Beijing 100084, China}

\author{Tong Zou}
\affiliation{State Key laboratory of quantum functional materials, Department of Physics, and Guangdong Basic Research Center of Excellence for Quantum Science, Southern University of Science and Technology (SUSTech), Shenzhen 518055, China}

\author{Ruiyue Chu}
\affiliation{Key Laboratory of Advanced Materials (MOE), School of Materials Science and Engineering, Tsinghua University, Beijing 100084, China}

\author{Gui Wang}
\author{Kaiwen Shen}
\author{Zhiwei Zhang}
\author{Fang Chen}
\author{Yingfen Wei}
\author{Hao Jiang}
\author{Xumeng Zhang}
\author{Ming Wang}
\affiliation{Frontier Institute of Chip and System, State Key Laboratory of Integrated Chips and Systems, Zhangjiang Fudan International Innovation Center, Fudan University, Shanghai 200433, China}

\author{Liyang Liao}
\affiliation{Shanghai Institute of Microsystem and Information Technology, ShanghaiTech University, Shanghai 201210, China}

\author{Jingli Wang}
\affiliation{Frontier Institute of Chip and System, State Key Laboratory of Integrated Chips and Systems, Zhangjiang Fudan International Innovation Center, Fudan University, Shanghai 200433, China}

\author{Yijun Yu}
\affiliation{Department of Physics, Fudan University, Shanghai 200433, China}

\author{Wu Shi}
\affiliation{State Key Laboratory of Surface Physics and Institute for Nanoelectronic Devices and Quantum Computing, Fudan University, Shanghai 200433, China}

\author{Qi Yao}
\affiliation{Shanghai Institute of Microsystem and Information Technology, ShanghaiTech University, Shanghai 201210, China}

\author{Xiaobing Chen}
\affiliation{Quantum Science Center of Guangdong-Hong Kong-Macao Greater Bay Area (Guangdong), Shenzhen 518045, China}

\author{Qihang Liu}
\affiliation{State Key laboratory of quantum functional materials, Department of Physics, and Guangdong Basic Research Center of Excellence for Quantum Science, Southern University of Science and Technology (SUSTech), Shenzhen 518055, China}
\affiliation{Quantum Science Center of Guangdong-Hong Kong-Macao Greater Bay Area (Guangdong), Shenzhen 518045, China}

\author{Yizheng Wu}
\affiliation{Department of Physics, Fudan University, Shanghai 200433, China}

\author{Yuhang Li}
\email{liyuhang@nankai.edu.cn}
\affiliation{School of Physics, Nankai University, Tianjin 300071, China}

\author{Xufeng Kou}
\email{kouxf@shanghaitech.edu.cn}
\affiliation{Shanghai Institute of Microsystem and Information Technology, ShanghaiTech University, Shanghai 201210, China}

\author{Cheng Song}
\email{songcheng@mail.tsinghua.edu.cn}
\affiliation{Key Laboratory of Advanced Materials (MOE), School of Materials Science and Engineering, Tsinghua University, Beijing 100084, China}

\author{Xianzhe Chen}
\email{chenxianzhe@fudan.edu.cn}
\affiliation{Frontier Institute of Chip and System, State Key Laboratory of Integrated Chips and Systems, Zhangjiang Fudan International Innovation Center, Fudan University, Shanghai 200433, China}

\author{Qi Liu}
\affiliation{Frontier Institute of Chip and System, State Key Laboratory of Integrated Chips and Systems, Zhangjiang Fudan International Innovation Center, Fudan University, Shanghai 200433, China}

\author{Xincheng Xie}
\affiliation{Interdisciplinary Center for Theoretical Physics and Information Sciences (ICTPIS), Fudan University, Shanghai 200433, China}
\affiliation{International Center for Quantum Materials, Peking University, Beijing 100871, China}
\affiliation{Hefei National Laboratory, Hefei 230088, China}

\author{Ming Liu}
\affiliation{Frontier Institute of Chip and System, State Key Laboratory of Integrated Chips and Systems, Zhangjiang Fudan International Innovation Center, Fudan University, Shanghai 200433, China}

\begin{abstract}
Altermagnetic materials, characterized by vanishing net magnetization and momentum-dependent spin splitting, provide a unique platform for realizing unconventional superconducting states without the detrimental stray fields. Here, we demonstrate long-distance Josephson supercurrent transport through epitaxial thin films of the strained altermagnet CrSb. We observe a dissipationless supercurrent across a lateral spacer distance of 150~nm in a CrSb-based Josephson junction, exceeding the conventional spin-singlet coherence length by nearly two orders of magnitude. This supercurrent is further accompanied by a well-defined magnetic-field-induced quantum interference pattern, providing direct evidence for genuine Josephson coupling. Our results reveal altermagnets as a promising route toward generating long-distance supercurrents without macroscopic magnetic stray fields, opening new opportunities for superconducting spintronic applications.
\end{abstract}

\maketitle

The interplay between magnetism and superconductivity gives rise to many emergent states~\cite{Buzdin2005Prominity,linder2015superconducting}, such as spin-polarized triplet Cooper pairs and finite-momentum Fulde–Ferrell–Larkin–Ovchinnikov states, holding great potential in dissipationless superconducting electronics~\cite{acharjee2026unconventional,bergeret2005odd,birge2018spin,robinson2010controlled,eschrig2015spin,birge2018spin}. Conventionally, the typical strategies to create these novel states rely on the fabricated heterostructure of a superconductor in proximity to a ferromagnet~\cite{Buzdin2005Prominity,golubov2004current,yamashita2005superconducting,jeon2021long,jeon2023chiral,zhang2025observation,PhysRevLett.86.4096,martinez2016amplitude}. These approaches inevitably bring in some critical limitations including  complex multilayer architectures, stringent requirements for nanoscale magnetization alignment, and stray-field-induced interference and vortex nucleation in adjacent superconductor~\cite{golubov2004current,yamashita2005superconducting,jeon2021long,jeon2023chiral,zhang2025observation,PhysRevLett.86.4096,martinez2016amplitude}.

Recently, an emerging frontier dubbed altermagnets provides a new platform to engineering these exotic states. In contrast to conventional ferromagnets, altermagnets possess zero net magnetization while maintaining momentum-dependent spin splitting arising from combined crystal symmetry operations (rotation or mirror) and time-reversal symmetry~\cite{PhysRevX.12.031042,PhysRevX.12.040501,song2025altermagnets,jungwirth2026symmetry,fender2025altermagnetism,krempasky2024altermagnetic,zhou2025manipulation,chen2025unconventional}. The absence of macroscopic magnetization substantially suppresses stray magnetic fields, while the intrinsic momentum-dependent spin polarization provides a natural route toward unconventional superconducting correlations analogous to ferromagnets. Theoretically, altermagnets have been predicted to host many novel superconducting phenomena including perfect superconducting diode effect~\cite{chakraborty2025perfect}, the finite momentum Fulde–Ferrell–Larkin–Ovchinnikov pairing~\cite{zhang_finite-momentum_2024,PhysRevResearch.5.043171}, spin-polarized triplet Cooper pairs~\cite{52wh-1z5y,PhysRevB.108.184505,PhysRevB.110.L220503,g4dl-1ff2}, $0-\pi$ phase oscillation~\cite{PhysRevLett.131.076003}, \textit{et.al.} Nevertheless, despite a very recent observation of a proximity-induced supercurrent along the surface of altermagnet CrSb in In-CrSb-In junctions~\cite{esin2026josephson}, direct experimental demonstration of the supercurrent in an altermagnet Josephson junction still remains scarce. 

In this Letter, we demonstrate the observation of a long-distance supercurrent in an altermagnet Josephson junction with strained CrSb sandwiched between two $s$-wave superconducting Nb leads. We observe dissipationless superconducting transport across a CrSb spacer with an edge-to-edge separation of up to 150~nm, substantially exceeding the characteristic coherence length expected for conventional spin-singlet correlations in a strongly magnetic environment. This supercurrent exhibits a pronounced temperature dependence and a characteristic Fraunhofer magnetic-field interference pattern, providing unambiguous evidence for its Josephson origin. We further discuss a possible microscopic conversion mechanism in which the momentum-dependent spin splitting of CrSb generates spin-mixed triplet correlation $|10\rangle$. Subsequently, the emergent spin-orbit coupling under strain-induced crystalline symmetry breaking, which is confirmed via the anomalous Hall resistance, further rotates this triplet correlation $|10\rangle$ toward equal-spin component $|11\rangle$ (or $|1\bar{1}\rangle$) triplet state, resulting in the long-distance supercurrent. Our results establish altermagnets as a promising platform for long-distance superconducting transport without the macroscopic stray fields associated with conventional ferromagnets, opening a new avenue for scalable superconducting spintronic devices.

\begin{figure}[t!]
    \centering
    \includegraphics[width=0.4\textwidth]{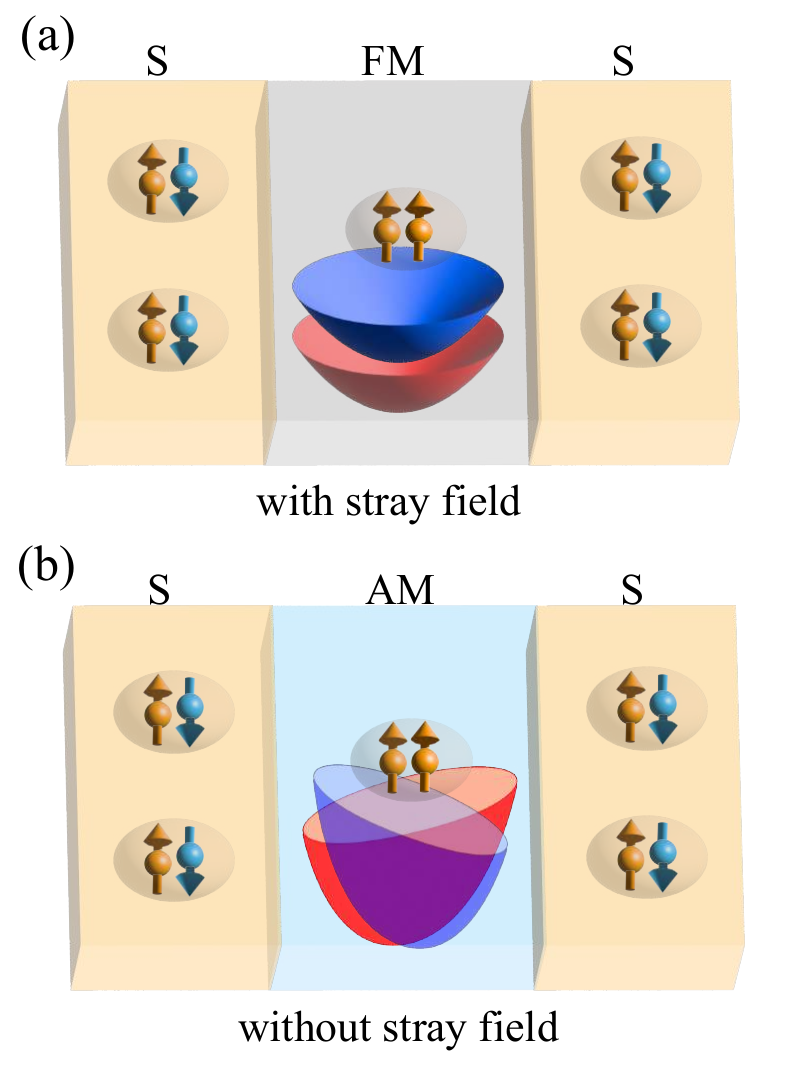}
    \caption{Schematics of (a) ferromagnetic and (b) altermagnetic Josephson junctions. Spin-split band structures are illustrated in each panel, respectively. Both systems can support spin-triplet pairing. Inside the $s$-wave superconducting leads, only spin-singlet Cooper pairs exist, which can be converted into spin triplets via spin-mixing and spin-rotation effects.
    }
    \label{fig:1}
\end{figure}

To clarify the fundamental difference between conventional ferromagnetic and altermagnetic Josephson junctions, figure~\ref{fig:1} compares their magnetic configurations and corresponding spin structures. As illustrated in Fig.~\ref{fig:1}(a), in conventional ferromagnetic Josephson junctions, the spontaneous magnetization produces macroscopic stray fields, which can induce screening currents and Abrikosov vortex formation in the neighboring superconducting electrodes~\cite{golod2019two}. These additional magnetic-field effects introduce phase inhomogeneity and complicate the identification of intrinsic triplet-mediated Josephson transport. In contrast, as shown in Fig.~\ref{fig:1}(b), an altermagnet exhibits compensated local spin polarization with vanishing net magnetization. The absence of macroscopic stray fields, combined with symmetry-protected momentum-dependent spin splitting, provides an ideal platform for exploring unconventional superconducting proximity effects free from the magnetic-field background associated with ferromagnets~\cite{PhysRevLett.86.4096}.

\begin{figure*}[t!]
    \centering
    \includegraphics[width=0.98\textwidth]{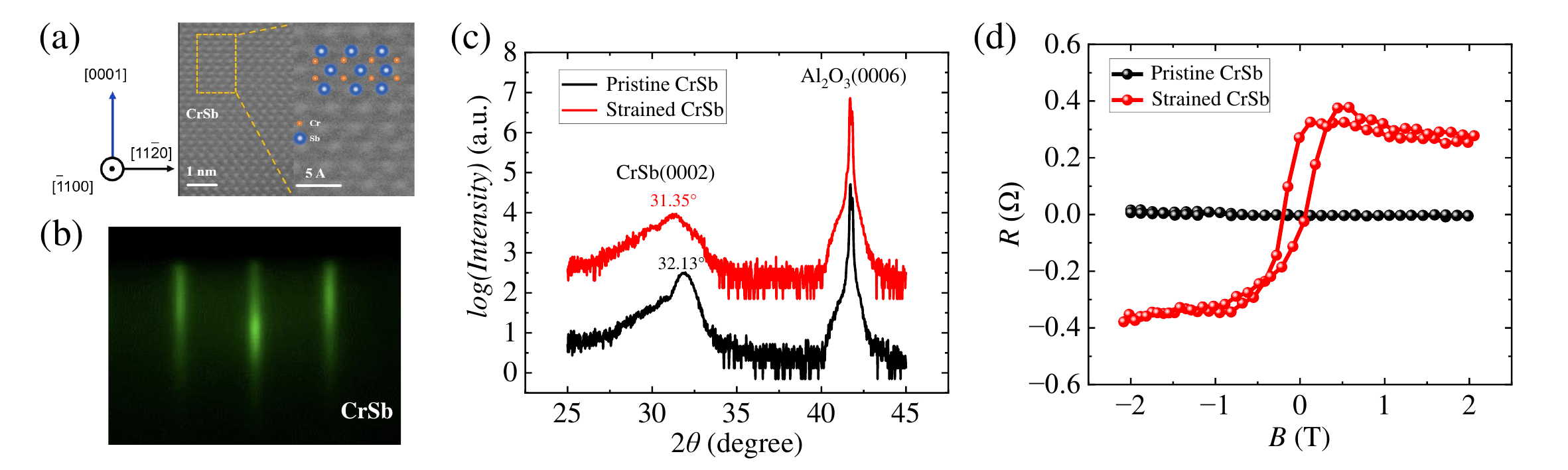}
    \caption{
    (a) Atomic-scale characterization of the CrSb crystal lattice synthesized via molecular beam epitaxy (MBE). 
    (b) Reflection high-energy electron diffraction (RHEED) pattern of the CrSb film, exhibiting sharp and streaky features indicative of smooth, two-dimensional epitaxial growth.
    (c) X-ray diffraction (XRD) patterns of pristine (black) and strained (red) CrSb films. The shift in the CrSb (0002) diffraction peak from $32.13^\circ$ to $31.35^\circ$ confirms the lattice expansion induced by strain.
    (d) Hall resistance $R_{xy}$ as a function of the external magnetic field $B$ for pristine (black spheres) and strained (red spheres) CrSb samples, demonstrating the modified magnetic response under strain.
    Here, the thickness of the CrSb sample is 10~nm.
    }
    \label{fig:2}
\end{figure*}

CrSb crystallizes in a hexagonal structure with the $P6_3/mmc$ lattice group~\cite{ding2024large}. The localized moments of neighboring Cr atoms are arranged antiparallelly, forming an A-type antiferromagnetic configuration in real space. CrSb has been identified as a $g$-wave altermagnet~\cite{reimers2024direct}, where the magnetic order produces momentum-dependent spin splitting while maintaining zero net magnetization. To establish the structural foundation for realizing high-quality Josephson devices, we systematically characterize molecular-beam-epitaxy (MBE)-grown CrSb thin films deposited on Al\textsubscript{2}O\textsubscript{3} substrates. Atomic-scale characterization in Fig.~\ref{fig:2}(a) further confirms the well-ordered crystalline lattice of CrSb, demonstrating the high structural quality of the epitaxial films enabled by MBE growth. This is further confirmed by the sharp and streaky features of the Reflection high-energy electron diffraction (RHEED) pattern in Fig.~\ref{fig:2}(b).

Besides, we further investigate CrSb films grown on Al${}_2$O${}_3$ substrates, where lattice mismatch between them introduces a uniform yet substantial strain into the CrSb lattice. As shown in Fig.~\ref{fig:2}(c), X-ray diffraction (XRD) measurements reveal distinct diffraction peaks corresponding to the pristine CrSb, which shift from 32.13$^\circ$ to 31.35$^\circ$ under strain induced lattice expansion.
This peak shift indicates that the lattice constant along the $c$-axis increases from $\sim0.2784$~nm to $\sim0.2851$~nm calculated using Bragg's law, leading to a relatively strong lattice expansion of $\delta c/c_{initial}\approx2.4\%$. This strain engineering provides an effective approach to modify the crystal symmetry and spin-dependent electronic structure of CrSb while preserving its compensated antiferromagnetic order~\cite{zhou2025manipulation}. It is worth noting that although the XRD measurement only reflects the lattice distortion along $c$-axis, the lateral strain is also possible. This is clearly evidenced by the anomalous Hall measurement under an applied magnetic field in Fig.~\ref{fig:2}(d)~\cite{zhou2025manipulation}. In pristine CrSb, the anomalous Hall resistance $R_{xy}$ vanishes completely, consistent with the compensated nature of the altermagnetic state. In contrast, the strained CrSb exhibits a finite anomalous Hall response that qualitatively follows the magnetic hysteresis behavior, providing evidence for strain-induced symmetry breaking and the emergence of spin-orbit coupling effect~\cite{Nagaosa2010}. Since the lattice strain along $c$-axis solely prohibits this anomalous Hall effect~\cite{zhou2025manipulation}, the non-vanishing $R_{xy}$ thus indicates the existence of lateral lattice strain and also the emergence of spin-orbit coupling. 

\begin{figure}[t!]
    \centering
    \includegraphics[width=0.45\textwidth]{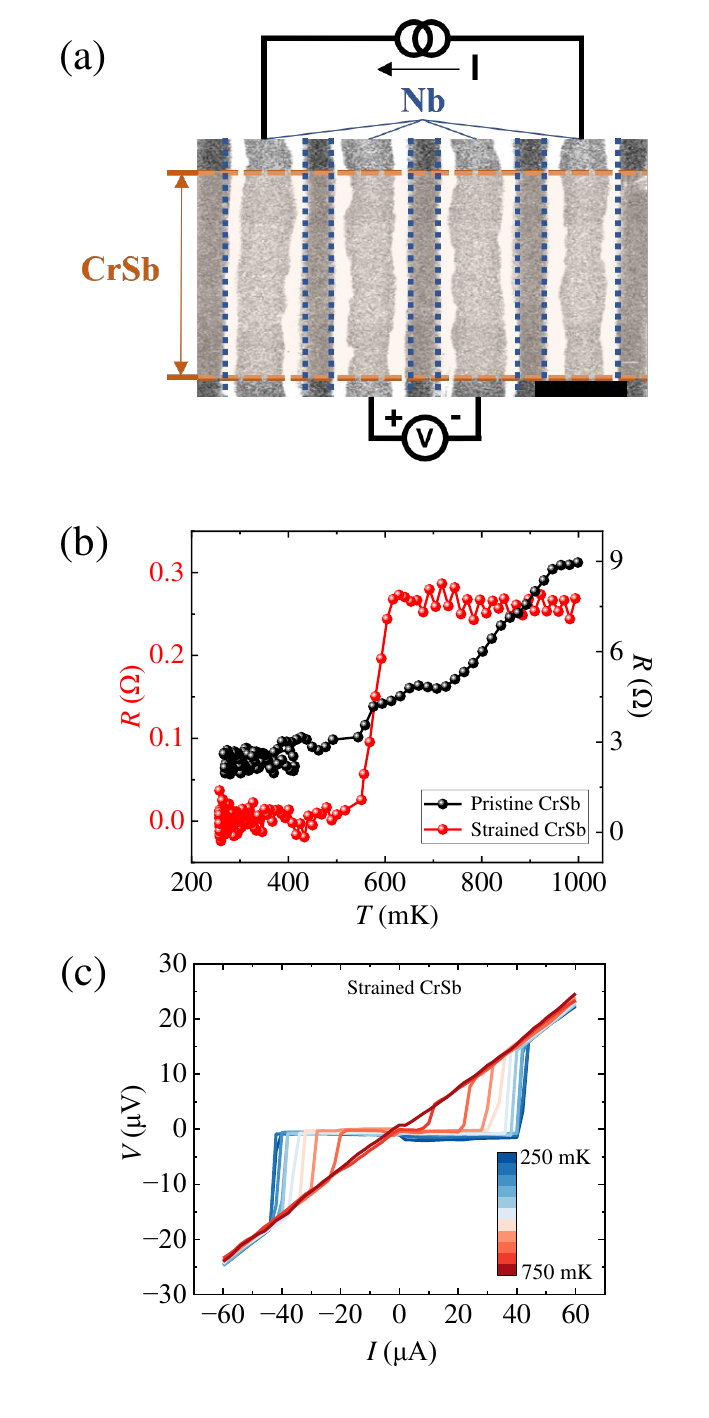}
    \caption{
    (a) SEM image of the CrSb Josephson junction device. Scale bar, $0.5~\mu$m. The four-terminal
resistive properties were acquired using a lock-in amplifier system (SR830).
    (b) Temperature dependence of the device resistance ($R-T$ curve). The junction exhibits a sharp superconducting transition, with the resistance dropping to zero at the temperature of $T\approx0.55$~K.
    (c) $I-V$ curve of the CrSb Josephson junction at various temperature ranging from 250~mK to 750~mK. The device becomes completely superconducting at the temperature $T\approx0.55$~K.
    }
    \label{fig:3}
\end{figure}

Leveraging this high quality CrSb, we fabricate lateral Josephson junctions consisting of a 10-nm-thick CrSb weak link contacted by 50-nm-thick Nb superconducting electrodes, with an edge-to-edge separation of 150~nm, as shown in Fig.~\ref{fig:3}(a). Four-terminal transport measurements are performed using a multichannel lock-in amplifier system with DC and AC biasing, at temperatures ranging from 250~mK to 750~mK and under out-of-plane magnetic fields up to $\pm$20~mT. In diffusive magnetic Josephson junctions, the propagation length of superconducting correlations is determined by the dominant pair-breaking mechanism. The characteristic coherence length can be estimated as~\cite{PhysRevLett.86.4096,jeon2021long}
\begin{align}\label{decay_length}
\xi=\sqrt{\frac{\hbar D}{4\pi k_BT_s}}, 
\end{align}
where $D$ is the diffusion coefficient of CrSb, $\hbar$ is the reduced Planck constant, $k_B$ is the Boltzmann constant and $T_s$ represents the critical temperature accounting for the main mechanism breaking the Cooper pairs~\cite{jeon2021long}. For conventional spin-singlet correlations penetrating into pristine CrSb, the dominant limitation arises from the strong exchange interaction associated with the magnetic Cr moments. The corresponding pair-breaking temperature is determined by the magnetic ordering temperature $T_N\approx 700$~K~\cite{zhou2025manipulation}. Using the diffusion coefficient $D\approx 0.01$~m${}^2$/s~\cite{long20263dbulkresolvedgwavemagnetic}, this strong magnetic pair breaking results in an estimated singlet coherence length of only $\xi_{\textrm{singlet}}\approx 3$~nm. In contrast, equal-spin triplet correlations are insensitive to the exchange-field-induced pair breaking because both electrons experience the same spin-dependent potential. Therefore, the propagation length of equal-spin triplet correlations is mainly limited by thermal dephasing rather than magnetic pair breaking. At $T=250$~mK, the resulting thermal coherence length reaches $\xi_{\textrm{triplet}}\approx 160$~nm, comparable to the fabricated junction length. Therefore, the observation of Josephson transport over a length scale far exceeding the singlet coherence limit strongly suggests the generation of long-distance spin-triplet Cooper pairs in the altermagnetic CrSb junction.
As mentioned above, this spin-triplet Cooper pair can be readily ascribed to the spin mixing and spin rotation effect~\cite{linder2015superconducting,eschrig2015spin,PhysRevLett.86.4096,khaire2010observation}. The momentum-dependent spin splitting of CrSb provides a natural spin-mixing mechanism that converts incoming singlet correlations from the Nb electrodes into the spin-triplet components $|10\rangle$. Subsequently, strain-induced spin-orbit coupling enables spin rotation from the triplet component $|10\rangle$ into equal-spin components $|11\rangle$ (or $|1\bar{1}\rangle$), persisting over a long distance about two orders of magnitude larger than the spin singlet Cooper pair and leading to a detectable Josephson supercurrent.

Figure~\ref{fig:3}(b) compares the temperature-dependent resistance ($R-T$ curve) of Josephson junctions fabricated from pristine (black spheres) and strained (red spheres) CrSb weak links with identical spacer lengths of $150$~nm. The strained CrSb device exhibits a pronounced superconducting transition below $T\approx0.55$~K, whereas the pristine CrSb junction does not show comparable zero-resistance transport. The absence of supercurrent in pristine CrSb with the same junction geometry rules out conventional metallic conduction or pinhole-mediated transport as the origin of the observed zero-resistance state. Additionally, in Fig.~\ref{fig:3}(c), the current–voltage ($I-V$) curves measured at different temperatures display a clear Josephson signature, featuring a zero-voltage branch at low bias currents and a temperature-dependent switching current. With increasing temperature, the extracted critical current $I_c$ decreases monotonically, consistent with the suppression of the superconducting order parameter.
This remarkable long-distance transport behavior provides strong evidence for the presence of spin-polarized triplet Cooper pairs, which are immune to exchange-field-induced pair breaking and can thus propagate over much longer distances than singlet pairs. 

\begin{figure}[t!]
    \centering
    \includegraphics[width=0.48\textwidth]{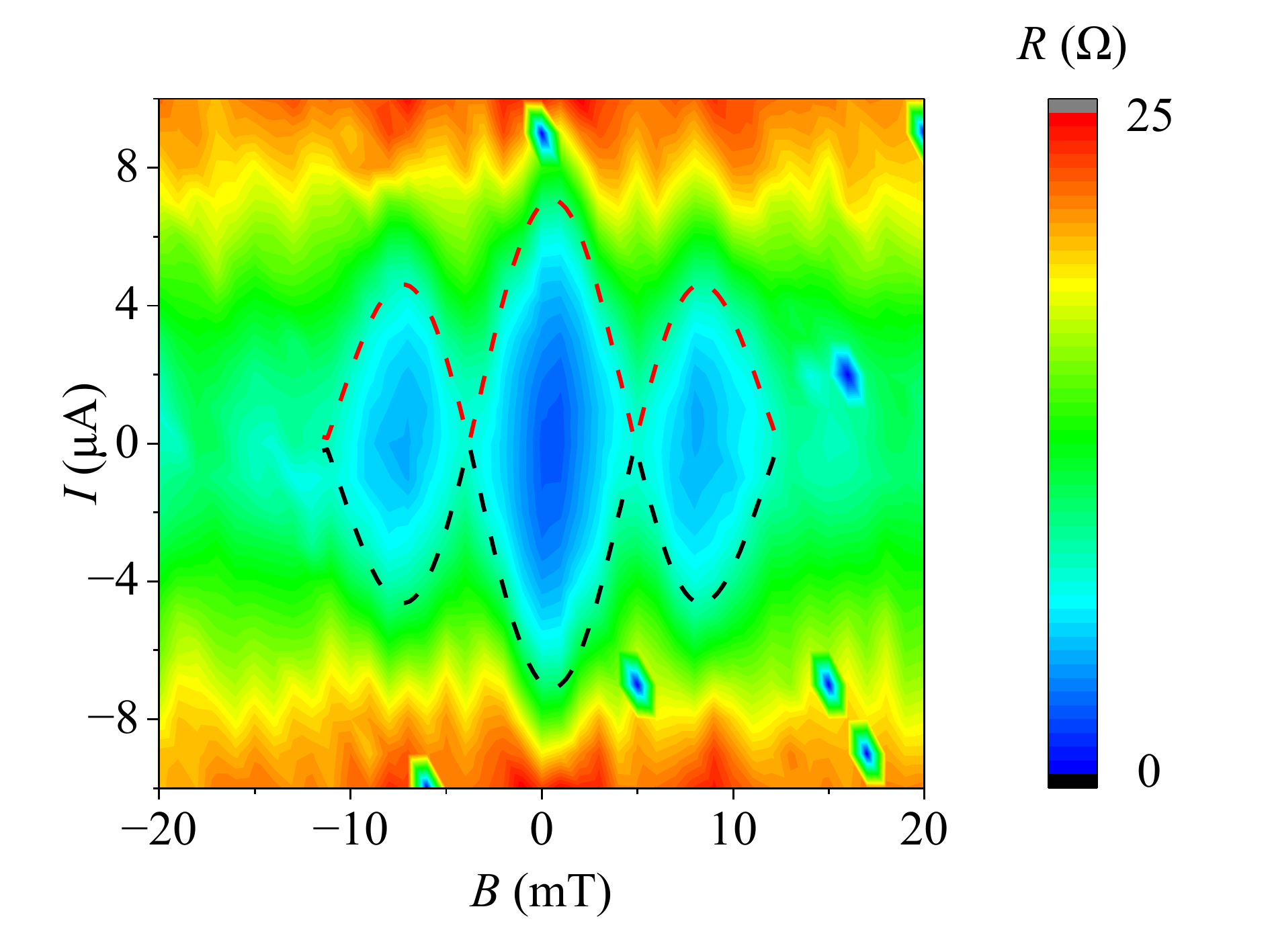}
    \caption{Color plot of the differential resistance ($dV/dI$) for the CrSb Josephson junction mapped against applied current $I$ and the out-of-plane magnetic field $B$ at 250~mK. The black and red dashed lines track the switching current $I_c$, highlighting a Fraunhofer interference pattern with enhanced first-order side lobes, indicative of highly extended edge supercurrents under spin-orbit coupling.
    }
    \label{fig:4}
\end{figure}

Figure~\ref{fig:4} presents the differential resistance map $R(B,I)$ of the strained CrSb junction as a function of external magnetic field $B$ and applied current $I$ at 250~mK. Despite the enhancement of the first lobes, the differential resistance exhibits a clear periodic modulation with a central maximum around $B\approx 0$ and nodes near $B\approx\pm4$~mT. Such magnetic-field-dependent interference is a characteristic fingerprint of spatial phase coherence in Josephson junctions~\cite{Dynes1971}. The enhancement of the first-order side lobes may be associated with a nonuniform current distribution arising from strain-induced spin-orbit coupling effects, which can redistribute the supercurrent toward the junction edges~\cite{li2022quantum,hart2014induced,wang2020evidence,veldhorst2012josephson}, in qualitative consistent with the observation of anomalous Hall effect in Fig.~\ref{fig:2}(d). For a spatially uniform Josephson junction, the critical current follows the conventional Fraunhofer relation~\cite{Dynes1971,li2022quantum}
\begin{align}\label{inter}
I_c(B)=I_c(0)\lvert \frac{\sin{\pi\Phi/\Phi_0}}{\pi\Phi/\Phi_0}\rvert
\end{align}
where $\Phi_0=2.07\times10^{-15}$~Wb is the magnetic flux quantum, $\Phi=BS$ is the total magnetic flux penetrating the central CrSb with $S$ its area. Therefore, the expected oscillation period is determined by the effective junction area. Using the junction geometry, we obtain an expected oscillation period of $\Delta B\approx8$~mT, consistent with the experimentally observed interference pattern in Fig.~\ref{fig:4}.

The microscopic origin of the long-distance Josephson transport in strained CrSb can be ascribed to the interplay between altermagnetic spin splitting and strain-induced spin-orbit coupling~\cite{linder2015superconducting,eschrig2015spin,PhysRevLett.86.4096,khaire2010observation}. The momentum-dependent spin splitting inherent to the $g$-wave altermagnetic order acts as an effective spin-mixing field, converting injected singlet Cooper pairs $|00\rangle$ from the Nb electrodes into $|10\rangle$ triplet correlations. The strain-induced symmetry breaking further activates spin-orbit coupling effects, which rotate the $|10\rangle$ triplet component into equal-spin triplet states $|11\rangle$ or $|1\bar{1}\rangle$, making the Cooper-pairs immune to the magnetic moments. This mechanism is conceptually related to the Bergeret–Volkov–Efetov theory developed for superconductor/ferromagnet/superconductor junctions~\cite{PhysRevLett.86.4096}, while providing a crucial advantage, where the compensated altermagnetic order generates no macroscopic stray field, thereby avoiding vortex-induced decoherence. On the other hand, since both space-inversion and time-reversal symmetries are simultaneously broken in altermagnetic CrSb, the emergence of a finite-momentum Fulde–Ferrell–Larkin–Ovchinnikov (FFLO) state is, in principle, feasible~\cite{zhang_finite-momentum_2024,PhysRevResearch.5.043171}. This could give rise to an unexplored superconducting diode effect~\cite{esin2026josephson}. Consequently, the supercurrent observed in this work provides a strong impetus for further investigations into altermagnetic superconductivity.

In conclusion, we have experimentally demonstrated long-distance Josephson supercurrents through the altermagnet CrSb over distances exceeding the conventional singlet coherence length by approximately two orders of magnitude. The combination of zero-resistance transport, magnetic-field-induced Josephson interference, and strain-dependent symmetry breaking provides possible evidence for unconventional triplet-mediated superconducting transport in altermagnets.
By replacing ferromagnetic interlayers with altermagnets, our approach removes the outstanding obstacles of stray fields and complex multilayer engineering in superconducting spintronics. 
Our findings establish altermagnets as a new material platform for scalable superconducting spintronics, enabling future exploration of phase-controllable Josephson devices, superconducting diodes, and triplet-based superconducting quantum interference technologies.

We acknowledge fruitful discussion with T. Zhou. This work is supported by the National Natural Science Foundation of China (Grants No. 12404056), the Tsinghua University Initiative Scientific Research Program.Yu-Hang Li also acknowledges the financial support from the Natural Science Foundation of Tianjin, China (Grant No. 24JCQNJC01910), the Fundamental Research Funds for the Central Universities (Grant No. 010-63263112).

\bibliography{bibliography}

@article{Buzdin2005Prominity,
  title = {Proximity effects in superconductor-ferromagnet heterostructures},
  author = {Buzdin, A. I.},
  journal = {Rev. Mod. Phys.},
  volume = {77},
  issue = {3},
  pages = {935--976},
  numpages = {0},
  year = {2005},
  month = {Sep},
  publisher = {American Physical Society},
  doi = {10.1103/RevModPhys.77.935}
}

@article{PhysRevX.12.031042,
  title = {Beyond Conventional Ferromagnetism and Antiferromagnetism: A Phase with Nonrelativistic Spin and Crystal Rotation Symmetry},
  author = {\ifmmode \check{S}\else \v{S}\fi{}mejkal, Libor and Sinova, Jairo and Jungwirth, Tomas},
  journal = {Phys. Rev. X},
  volume = {12},
  issue = {3},
  pages = {031042},
  numpages = {16},
  year = {2022},
  month = {Sep},
  publisher = {American Physical Society},
  doi = {10.1103/PhysRevX.12.031042},
}

@article{PhysRevX.12.040501,
  title = {Emerging Research Landscape of Altermagnetism},
  author = {\ifmmode \check{S}\else \v{S}\fi{}mejkal, Libor and Sinova, Jairo and Jungwirth, Tomas},
  journal = {Phys. Rev. X},
  volume = {12},
  issue = {4},
  pages = {040501},
  numpages = {27},
  year = {2022},
  month = {Dec},
  publisher = {American Physical Society},
  doi = {10.1103/PhysRevX.12.040501},
}

@article{song2025altermagnets,
  title={Altermagnets as a new class of functional materials},
  author={Song, Cheng and Bai, Hua and Zhou, Zhiyuan and Han, Lei and Reichlova, Helena and Dil, J Hugo and Liu, Junwei and Chen, Xianzhe and Pan, Feng},
  journal={Nat. Rev. Mater.},
  volume={10},
  number={6},
  pages={473--485},
  year={2025},
  publisher={Nature Publishing Group UK London}
}

@article{jungwirth2026symmetry,
  title={Symmetry, microscopy and spectroscopy signatures of altermagnetism},
  author={Jungwirth, Tomas and Sinova, Jairo and Fernandes, Rafael M and Liu, Qihang and Watanabe, Hikaru and Murakami, Shuichi and Nakatsuji, Satoru and {\v{S}}mejkal, Libor},
  journal={Nature},
  volume={649},
  number={8098},
  pages={837--847},
  year={2026},
  publisher={Nature Publishing Group UK London}
}

@article{fender2025altermagnetism,
  title={Altermagnetism: A chemical perspective},
  author={Fender, Shannon S and Gonzalez, Oscar and Bediako, D Kwabena},
  journal={J. Am. Chem. Soc.},
  volume={147},
  number={3},
  pages={2257--2274},
  year={2025},
  publisher={ACS Publications}
}

@article{krempasky2024altermagnetic,
  title={Altermagnetic lifting of Kramers spin degeneracy},
  author={Krempask{\`y}, Juraj and {\v{S}}mejkal, L and D’souza, SW and Hajlaoui, M and Springholz, G and Uhl{\'\i}{\v{r}}ov{\'a}, K and Alarab, F and Constantinou, PC and Strocov, V and Usanov, D and others},
  journal={Nature},
  volume={626},
  number={7999},
  pages={517--522},
  year={2024},
  publisher={Nature Publishing Group UK London}
}

@article{zhou2025manipulation,
  title={{Manipulation of the altermagnetic order in CrSb via crystal symmetry}},
  author={Zhou, Zhiyuan and Cheng, Xingkai and Hu, Mengli and Chu, Ruiyue and Bai, Hua and Han, Lei and Liu, Junwei and Pan, Feng and Song, Cheng},
  journal={Nature},
  volume={638},
  number={8051},
  pages={645--650},
  year={2025},
  publisher={Nature Publishing Group UK London}
}

@article{chen2025unconventional,
  title={Unconventional magnons in collinear magnets dictated by spin space groups},
  author={Chen, Xiaobing and Liu, Yuntian and Liu, Pengfei and Yu, Yutong and Ren, Jun and Li, Jiayu and Zhang, Ao and Liu, Qihang},
  journal={Nature},
  volume={640},
  number={8058},
  pages={349--354},
  year={2025},
  publisher={Nature Publishing Group UK London}
}

@article{long20263dbulkresolvedgwavemagnetic,
  author        = {Long, Mengmeng and Weinberger, Theodore I. and Wu, Zheyu and Hansen, Mads F. and Tao, Ran and Shrestha, Mridul and Graf, Dave and Skourski, Yurii and Grosche, F. Malte and Eaton, Alexander G.},
  title         = {{3D bulk-resolved $g$-wave magnetic order parameter symmetry in the metallic altermagnet CrSb}},
  journal       = {arXiv preprint arXiv:2601.14526},
  year          = {2026},
  eprint        = {2601.14526},
  archivePrefix = {arXiv},
  primaryClass  = {cond-mat.mtrl-sci}
}

@article{Nagaosa2010,
  title = {{Anomalous Hall effect}},
  author = {Nagaosa, Naoto and Sinova, Jairo and Onoda, Shigeki and MacDonald, A. H. and Ong, N. P.},
  journal = {Rev. Mod. Phys.},
  volume = {82},
  issue = {2},
  pages = {1539--1592},
  numpages = {0},
  year = {2010},
  month = {May},
  publisher = {American Physical Society}
}

@article{reimers2024direct,
  title={{Direct observation of altermagnetic band splitting in CrSb thin films}},
  author={Reimers, Sonka and Odenbreit, Lukas and {\v{S}}mejkal, Libor and Strocov, Vladimir N and Constantinou, Procopios and Hellenes, Anna B and Jaeschke Ubiergo, Rodrigo and Campos, Warlley H and Bharadwaj, Venkata K and Chakraborty, Atasi and others},
  journal={Nat. Commun.},
  volume={15},
  number={1},
  pages={2116},
  year={2024},
  publisher={Nature Publishing Group UK London}
}

@article{ding2024large,
  title={{Large band splitting in $g$-wave altermagnet CrSb}},
  author={Ding, Jianyang and Jiang, Zhicheng and Chen, Xiuhua and Tao, Zicheng and Liu, Zhengtai and Li, Tongrui and Liu, Jishan and Sun, Jianping and Cheng, Jinguang and Liu, Jiayu and others},
  journal={Phys. Rev. Lett.},
  volume={133},
  number={20},
  pages={206401},
  year={2024},
  publisher={APS}
}

@article{linder2015superconducting,
  title={Superconducting spintronics},
  author={Linder, Jacob and Robinson, Jason WA},
  journal={Nat. Phys.},
  volume={11},
  number={4},
  pages={307--315},
  year={2015},
  publisher={Nature Publishing Group UK London}
}

@article{acharjee2026unconventional,
  title={Unconventional spin valve effect in altermagnets induced by Rashba spin orbit coupling and triplet superconductivity},
  author={Acharjee, Saumen and Dihingia, Aklanta and Sonowal, Nayanav and Kashyap, Abyoy Anan},
  journal={arXiv preprint arXiv:2602.20838},
  year={2026}
}

@article{bergeret2005odd,
  title={Odd triplet superconductivity and related phenomena in superconductor-ferromagnet structures},
  author={Bergeret, FS and Volkov, Anatoly F and Efetov, Konstantin B},
  journal={Rev. Mod. Phys.},
  volume={77},
  number={4},
  pages={1321--1373},
  year={2005},
  publisher={APS}
}

@article{birge2018spin,
  title={Spin-triplet supercurrents in Josephson junctions containing strong ferromagnetic materials},
  author={Birge, Norman O},
  journal={Philos. Trans. R. Soc. A},
  volume={376},
  number={2125},
  year={2018},
  publisher={The Royal Society}
}

@article{robinson2010controlled,
  title={Controlled injection of spin-triplet supercurrents into a strong ferromagnet},
  author={Robinson, JWA and Witt, JDS and Blamire, MG},
  journal={Science},
  volume={329},
  number={5987},
  pages={59--61},
  year={2010},
  publisher={American Association for the Advancement of Science}
}

@article{eschrig2015spin,
  title={Spin-polarized supercurrents for spintronics: a review of current progress},
  author={Eschrig, Matthias},
  journal={Rep. Prog. Phys.},
  volume={78},
  number={10},
  pages={104501},
  year={2015},
  publisher={IOP Publishing}
}

@article{khaire2010observation,
  title={{Observation of spin-triplet superconductivity in Co-based Josephson junctions}},
  author={Khaire, Trupti S and Khasawneh, Mazin A and Pratt Jr, WP and Birge, Norman O},
  journal={Phys. Rev. Lett.},
  volume={104},
  number={13},
  pages={137002},
  year={2010},
  publisher={APS}
}

@article{golubov2004current,
  title={{The current-phase relation in Josephson junctions}},
  author={Golubov, Alexandre Avraamovitch and Kupriyanov, M Yu and Il’Ichev, E},
  journal={Rev. Mod. Phys.},
  volume={76},
  number={2},
  pages={411},
  year={2004},
  publisher={APS}
}

@article{yamashita2005superconducting,
  title={{Superconducting $\pi$ qubit with a ferromagnetic Josephson junction}},
  author={Yamashita, Taro and Tanikawa, Kazuo and Takahashi, Satoru and Maekawa, Sadamichi},
  journal={Phys. Rev. Lett.},
  volume={95},
  number={9},
  pages={097001},
  year={2005},
  publisher={APS}
}

@article{jeon2021long,
  title={{Long-range supercurrents through a chiral non-collinear antiferromagnet in lateral Josephson junctions}},
  author={Jeon, Kun-Rok and Hazra, Binoy Krishna and Cho, Kyungjune and Chakraborty, Anirban and Jeon, Jae-Chun and Han, Hyeon and Meyerheim, Holger L and Kontos, Takis and Parkin, Stuart SP},
  journal={Nat. Mater.},
  volume={20},
  number={10},
  pages={1358--1363},
  year={2021},
  publisher={Nature Publishing Group UK London}
}

@article{jeon2023chiral,
  title={{Chiral antiferromagnetic Josephson junctions as spin-triplet supercurrent spin valves and dc SQUIDs}},
  author={Jeon, Kun-Rok and Hazra, Binoy Krishna and Kim, Jae-Keun and Jeon, Jae-Chun and Han, Hyeon and Meyerheim, Holger L and Kontos, Takis and Cottet, Audrey and Parkin, Stuart SP},
  journal={Nat. Nanotechnol.},
  volume={18},
  number={7},
  pages={747--753},
  year={2023},
  publisher={Nature Publishing Group UK London}
}

@article{zhang2025observation,
  title={{Observation of edge supercurrent in topological antiferromagnet {MnBi$_2$Te$_4$}-based Josephson junctions}},
  author={Zhang, Enze and Sun, Zi-Ting and Jia, Zehao and Yang, Jinshan and Yan, Jingyi and Ai, Linfeng and Xie, Ying-Ming and Zhang, Yuda and Gao, Xue-Jian and Xu, Xian and others},
  journal={Sci. Adv.},
  volume={11},
  number={20},
  pages={eads8730},
  year={2025},
  publisher={American Association for the Advancement of Science}
}

@article{PhysRevLett.86.4096,
  title = {{Long-Range Proximity Effects in Superconductor-Ferromagnet Structures}},
  author = {Bergeret, F. S. and Volkov, A. F. and Efetov, K. B.},
  journal = {Phys. Rev. Lett.},
  volume = {86},
  issue = {18},
  pages = {4096--4099},
  numpages = {0},
  year = {2001},
  month = {Apr},
  publisher = {American Physical Society},
  doi = {10.1103/PhysRevLett.86.4096},
}

@article{martinez2016amplitude,
  title={{Amplitude control of the spin-triplet supercurrent in S/F/S Josephson junctions}},
  author={Martinez, William M and Pratt Jr, WP and Birge, Norman O},
  journal={Phys. Rev. Lett.},
  volume={116},
  number={7},
  pages={077001},
  year={2016},
  publisher={APS}
}

@article{chakraborty2025perfect,
  title = {Perfect Superconducting Diode Effect in Altermagnets},
  author = {Chakraborty, Debmalya and Black-Schaffer, Annica M.},
  journal = {Physical Review Letters},
  volume = {135},
  pages = {026001},
  year = {2025},
  month = {Jul},
  publisher = {American Physical Society},
  doi = {10.1103/PhysRevLett.135.026001}
}

@article{Dynes1971,
  title = {{Supercurrent Density Distribution in Josephson Junctions}},
  author = {Dynes, R. C. and Fulton, T. A.},
  journal = {Phys. Rev. B},
  volume = {3},
  issue = {9},
  pages = {3015--3023},
  numpages = {0},
  year = {1971},
  month = {May},
  publisher = {American Physical Society},
  doi = {10.1103/PhysRevB.3.3015}
}

@article{li2022quantum,
  title={{Quantum interference in a superconductor-{MnBi$_2$Te$_4$}-superconductor Josephson junction}},
  author={Li, Yu-Hang and Cheng, Ran},
  journal={Phys. Rev. Res.},
  volume={4},
  number={3},
  pages={033227},
  year={2022},
  publisher={APS}
}

@article{hart2014induced,
  title={{Induced superconductivity in the quantum spin Hall edge}},
  author={Hart, Sean and Ren, Hechen and Wagner, Timo and Leubner, Philipp and M{\"u}hlbauer, Mathias and Br{\"u}ne, Christoph and Buhmann, Hartmut and Molenkamp, Laurens W and Yacoby, Amir},
  journal={Nat. Phys.},
  volume={10},
  number={9},
  pages={638--643},
  year={2014},
  publisher={Nature Publishing Group UK London}
}

@article{wang2020evidence,
  title={{Evidence for an edge supercurrent in the Weyl superconductor MoTe${}_2$}},
  author={Wang, Wudi and Kim, Stephan and Liu, Minhao and Cevallos, FA and Cava, Robert J and Ong, NP},
  journal={Science},
  volume={368},
  number={6490},
  pages={534--537},
  year={2020},
  publisher={American Association for the Advancement of Science}
}

@article{veldhorst2012josephson,
  title={Josephson supercurrent through a topological insulator surface state},
  author={Veldhorst, Menno and Snelder, Marieke and Hoek, Marcel and Gang, Tian and Guduru, VK and Wang, XL and Zeitler, Uli and van der Wiel, Wilfred Gerard and Golubov, AA and Hilgenkamp, Hans and others},
  journal={Nat. Mater.},
  volume={11},
  number={5},
  pages={417--421},
  year={2012},
  publisher={Nature Publishing Group UK London}
}

@article{PhysRevB.108.184505,
  title = {Topological superconductivity in two-dimensional altermagnetic metals},
  author = {Zhu, Di and Zhuang, Zheng-Yang and Wu, Zhigang and Yan, Zhongbo},
  journal = {Phys. Rev. B},
  volume = {108},
  issue = {18},
  pages = {184505},
  numpages = {13},
  year = {2023},
  month = {Nov},
  publisher = {American Physical Society},
  doi = {10.1103/PhysRevB.108.184505}
}

@article{PhysRevB.110.L220503,
  title = {Unconventional superconductivity in altermagnets with spin-orbit coupling},
  author = {de Carvalho, Vanuildo S. and Freire, Hermann},
  journal = {Phys. Rev. B},
  volume = {110},
  issue = {22},
  pages = {L220503},
  numpages = {6},
  year = {2024},
  month = {Dec},
  publisher = {American Physical Society},
  doi = {10.1103/PhysRevB.110.L220503}
}

@article{g4dl-1ff2,
  title = {Phonon-mediated spin-polarized superconductivity in altermagnets},
  author = {Leraand, Kristoffer and M\ae{}land, Kristian and Sudb\o{}, Asle},
  journal = {Phys. Rev. B},
  volume = {112},
  issue = {10},
  pages = {104510},
  numpages = {12},
  year = {2025},
  month = {Sep},
  publisher = {American Physical Society},
  doi = {10.1103/g4dl-1ff2}
}

@article{52wh-1z5y,
  title = {Persistent Spin Currents in Superconducting Altermagnets},
  author = {Monkman, Kyle and Weng, Joan and Heinsdorf, Niclas and Nocera, Alberto and Barlas, Yafis and Franz, Marcel},
  journal = {Phys. Rev. X},
  volume = {16},
  issue = {1},
  pages = {011057},
  numpages = {21},
  year = {2026},
  month = {Mar},
  publisher = {American Physical Society},
  doi = {10.1103/52wh-1z5y},
  url = {https://link.aps.org/doi/10.1103/52wh-1z5y}
}

@article{zhang_finite-momentum_2024,
	title = {Finite-momentum {Cooper} pairing in proximitized altermagnets},
	volume = {15},
	issn = {2041-1723},
	doi = {10.1038/s41467-024-45951-3},
	number = {1},
	journal = {Nature Communications},
	author = {Zhang, Song-Bo and Hu, Lun-Hui and Neupert, Titus},
	month = feb,
	year = {2024},
	pages = {1801},
}

@article{PhysRevResearch.5.043171,
  title = {Fulde-Ferrell-Larkin-Ovchinnikov state induced by antiferromagnetic order in $\ensuremath{\kappa}$-type organic conductors},
  author = {Sumita, Shuntaro and Naka, Makoto and Seo, Hitoshi},
  journal = {Phys. Rev. Res.},
  volume = {5},
  issue = {4},
  pages = {043171},
  numpages = {17},
  year = {2023},
  month = {Nov},
  publisher = {American Physical Society},
  doi = {10.1103/PhysRevResearch.5.043171},
  url = {https://link.aps.org/doi/10.1103/PhysRevResearch.5.043171}
}

@article{PhysRevLett.131.076003,
  title = {dc Josephson Effect in Altermagnets},
  author = {Ouassou, Jabir Ali and Brataas, Arne and Linder, Jacob},
  journal = {Phys. Rev. Lett.},
  volume = {131},
  issue = {7},
  pages = {076003},
  numpages = {6},
  year = {2023},
  month = {Aug},
  publisher = {American Physical Society},
  doi = {10.1103/PhysRevLett.131.076003},
  url = {https://link.aps.org/doi/10.1103/PhysRevLett.131.076003}
}

@article{golod2019two,
  title={Two mechanisms of Josephson phase shift generation by an Abrikosov vortex},
  author={Golod, Taras and Pagliero, Alessandro and Krasnov, Vladimir M},
  journal={Phys. Rev. B},
  volume={100},
  number={17},
  pages={174511},
  year={2019},
  publisher={APS}
}

@article{esin2026josephson,
  author  = {Esin, V. D. and Kazmin, D. Yu. and Barash, Yu. S. and Timonina, A. V. and Kolesnikov, N. N. and Deviatov, E. V.},
  title   = {Josephson diode and spin-valve effects on the surface of altermagnet CrSb},
  journal = {JETP Lett.},
  volume  = {123},
  pages   = {556},
  year    = {2026},
  doi     = {10.1134/S0021364026600667}
}
\end{document}